# A novel activity choice and scheduling model to model activity schedules of customers and staff in Dutch restaurants

**Martijn Sparnaaij, MSc. (Corresponding Author)**
Transport & Planning – Delft University of Technology
Stevinweg 1, 2628 CN, Delft, The Netherlands
Email: m.sparnaaij@tudelft.nl

**Dr. Yufei Yuan**
Transport & Planning – Delft University of Technology
Stevinweg 1, 2628 CN, Delft, The Netherlands
Email: y.yuan@tudelft.nl

**Dr. Winnie Daamen**
Transport & Planning – Delft University of Technology
Stevinweg 1, 2628 CN, Delft, The Netherlands
Email: w.daamen@tudelft.nl

**Dr. Dorine C. Duives**
Transport & Planning – Delft University of Technology
Stevinweg 1, 2628 CN, Delft, The Netherlands
Email: d.c.duives@tudelft.nl

(Note: A paper describing a completely overhauled version of the activity model is currently under review)

**ABSTRACT**
The Covid-19 pandemic has had a large impact on the world. The virus spreads especially easily among people in indoor spaces such as restaurants. Hence, tools that can assess if and how different restaurant settings can impact the potential spread of the virus are of high value. The novel activity choice and scheduling model for restaurants, presented in this paper, is a key part of a larger model which combines pedestrian dynamics modelling and epidemiological models to achieve exactly this.
This novel activity model can produce activity schedules for both customers and staff at a restaurant for a variety of restaurant settings. A key feature of the model is that it requires little user inputs to do so which is important as the intended users are restaurant owners. These owners have neither the time nor the expertise to deal with complex input.
Tests show that the model can produce face-valid activity schedules for both staff and customers for a variety of restaurant settings. The tests also show that different restaurant settings lead to distinctly different contact patterns between people in a restaurant. As such, the model can provide valuable insights into how a restaurant setting relates to the risk of virus transmission. This is especially the case when it is combined with data about the virus. Hence, it shows the high relevance of pedestrian dynamics modelling in these pandemic times and especially the relevance of activity choice and scheduling models.

# INTRODUCTION

Since its introduction at the end of 2019, the SARS-CoV-2 virus has had a major impact on the world, with over 180 million cases and almost 4 million deaths world-wide [1]. This has prompted governments worldwide to take a wide array of measures to curb the spread of this virus. Next to vaccination, which in parts of the world is well underway, governments have resorted to closing down many indoor places, such as schools, sports facilities, theatres and restaurants, as most infections occur indoors [2].

These closures have had major economic impacts, as [3] for example shows in the case of restaurants. Hence, it is desirable to reopen these places as quickly as possible. A group of measures, collectively known as non-pharmaceutical interventions (NPI's), are the primary tools of countries worldwide to reduce virus spread in indoor environments. Well-known examples of these NPI's are physical distancing regulations and facemasks. Therefore, it is of major importance to understand how these NPI's affect the spread of the virus in these indoor environments as this provides information on how to safely reopen places such as restaurants and to assist in a more targeted implementation of these NPI's.

Pedestrian dynamics models can support the NPI's impact analyses, as for example shown by [4]. Most of these modelling attempts directly translate the simulated operational movement dynamics to infections by means of a set of thresholds (e.g. all people that spend more than 15 minutes within 1.5 meters of an infectious individual). Yet, given that virus transmission is not a deterministic process, these endeavours under- or overestimate transmission risk in indoor and outdoor spaces.

Delft University of Technology (active mode lab) and Wageningen University & Research (epidemiology) have jointly developed the SamenSlimOpen application (SSOapp) to assess the impact of NPI's on the virus transmission risk. The SSOapp integrates the pedestrian dynamics modelling state-of-the-art with innovative virus spreading models [5]. The strength of this new SSOapp is that it explicitly models virus spread as a result of operational movement dynamics and is specifically developed for indoor spaces. This allows for a more specific and comprehensive analyses of virus transmission risks for indoor spaces.

To model the pedestrian dynamics in indoor spaces one needs to take into account several levels of pedestrian behaviour. Commonly, these are classified according to the strategic, tactical and operational levels [6]. The most important processes per level are: Activity choice on the strategic level; activity scheduling and route choice on the tactical level and short-term movement dynamics on the operational level. Within the SSOapp, NOMAD [7] is used as the basis to model the pedestrian behaviour, which can model the operational behaviour and the route choice of pedestrian crowds. However, activity choice and activity scheduling are not implemented systematically, but essential model inputs to derive valid operational movement dynamics and correctly identify virus transmission risks.

The lack of an integrated activity choice and scheduling model represents a problem for the SSOapp for the following reason. The first application of the tool is informing restaurant owners in how different interventions can reduce the spread of the virus in their restaurants once they reopen. And, as restaurant owners neither have much experience or expertise in pedestrian and virus transmission modelling, nor much time to spend on setting up these simulations, the tool must be simple as well as require limited user input. Hence, we need an activity choice and scheduling model that alleviates the restaurant owners from having to provide the tool with the activity schedules for all customers and staff.

A review of the literature turned up a few models focussed at modelling activity choice and scheduling of customers and staff at restaurants [8, 9]. However, these papers 1) focus on optimization of customer satisfaction or limiting waiting times for a table instead of the number and duration of contact between people in a restaurant environment and 2) the model description and evaluation are very limited. A wider search in the literature for non-restaurant related models turned up more. For example, activity models focussed on the activity choice and scheduling of pedestrians in a shopping street using discrete choice models [10**]**. Or rule-based approaches when modelling the activity choice and scheduling behaviour within an airport context [11]. Lastly, also an approach based on Markov chains to model the choice and scheduling of people within an office environment has been presented [12].

Overall, the literature shows that activity choice and scheduling models for pedestrians have received little attention. However, the limited research that is available shows multiple methods to tackle the problem. Yet, no activity choice and scheduling model was found that allows us to model the activity choice and scheduling of customer and waiting staff in restaurants, in such a way that realistic estimates of contacts between different people can be obtained.

This paper aims to fill that gap by presenting a newly developed rule-based activity choice and scheduling model for the indoor spaces. In this paper, the next section provides a detailed description of the newly developed model. Next, we explore the capabilities of the new model by means of a set of tests. Afterwards, the results of the test are presented and discussed. Lastly, the main conclusions are presented, the strengths and limitations of the model are discussed and its implications for practice are presented.

## THE ACTIVITY CHOICE AND SCHEDULING MODEL FOR RESTAURANT CUSTOMERS AND WAITING STAFF

This section describes the new restaurant activity choice and scheduling model. First, the intended goal and use if the model will be explained, directly followed by the set of practical limitations faced by the researchers. Based on the goals and requirements and the insights from the previously discussed literature, the basic design of the model is presented. Afterwards, a detailed explanation of the model is provided.

### Goal activity choice and scheduling model

The intended goal of the model is to provide activity schedules for all customers and waiting staff in a restaurant. Together with the already implemented route choice and movement dynamics model, these schedules can provide a realistic estimate of the contacts between different people within a typical dine-in restaurant setting. The focus of this modelling endeavour is on the customer-customer contacts and the staff-customer contacts, rather than the overall capacity or operation of the spaces. Please note, our focus is not on quantifying staff-staff contacts, because staff members are under all circumstances very likely to be in close proximity of each other for extended durations and hence other measures, such as regular testing, are more applicable to control staff-staff transmissions.

### SSOapp application context

The context within which this model is intended to be applied is dine-in restaurants. The main features that distinguish typical dine-in restaurants from other types of restaurants, such as take-out, fast-food or buffet-styled restaurants, is that 1) people eat the ordered food in the restaurant and 2) waiting staff serves the food. Furthermore, the intended users of the model are restaurant owners, who neither have much experience or expertise in pedestrian and virus transmission modelling nor much time to spend on setting up these simulations.

### Structure activity choice and scheduling model

Figure 1 presents an overview of how the newly designed model integrates with the Nomad model. The model features two sub-models, one for the customers and one for the waiting staff. Each sub-model has a different approach and interacts differently with the Nomad model. The core difference between both sub-models is that the customer sub-model provides static activity schedules for the customers to the Nomad model at the start of the simulation. The staff sub-model, on the other hand, provides on-the-fly assignments to staff members assigning their next activity dynamically during the simulation. The next two subsections provide more detail about the design and design choices of both these sub-models.

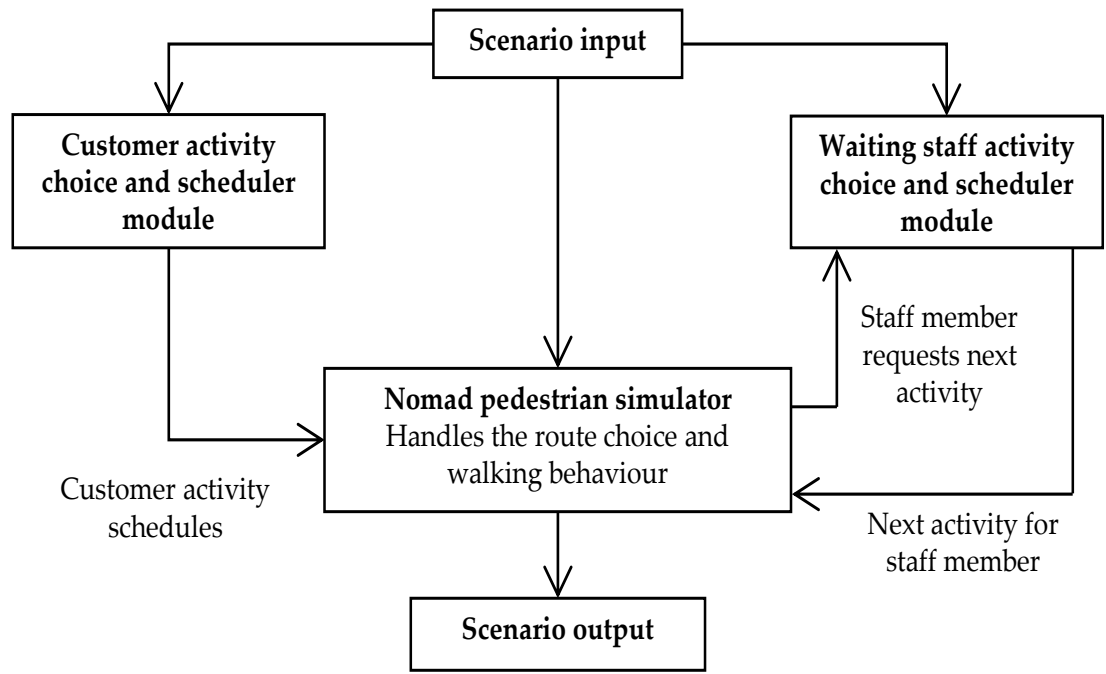

**Figure 1** Overview of the integration of the customer and waiting staff activity choice and scheduling model into the Nomad simulator

## Customer activity choice and scheduling model
The design and implementation of the customer sub-model will be explained in three steps. First, the input is discussed whereby the focus is upon the information that restaurant owners can realistically and easily provide. Second, the scheduling of visit of every group of customers is explained (i.e. when do they start their visit, at what table and when do they leave). And third, we show how the activity pattern of each individual member of a group is determined based on the input and the schedule of each group.

*Inputs*
Restaurant owners are the intended group of users of the model, therefore the inputs to the model should align with the inputs they can easily provide. Based on contacts with people in the restaurant industry and our own insights into what inputs are necessary, we selected the following inputs:
1. The restaurant layout
2. The time period that should be simulated
3. The demand pattern: This input divides the overall time period into smaller time slots and for each of those defines how many groups will visit the restaurant during that time.
4. The expected duration of the customer visits $t_{visit;expected}$

Restaurant owners can input these numbers into the SSOapp using a customer-friendly Graphical User Interface (GUI).

*Group scheduling*
Based on the input described above, the groups are scheduled. Yet, before the implementation is presented the expected behaviour is discussed. The scheduler is expected to schedule the groups such that:
- The groups start and end their visit within the time slot defined by the demand pattern
- Time slots can overlap. However time slots cannot have more groups than the number of table in the restaurant.

- If, during the time slot, no unoccupied table is available for at least the duration defined by the expected visit duration, a group that did reserve a spot, is not scheduled.
- A group should be able to occupy their assigned table for at least a time equal to the expected visit duration although they can stay shorter if they choose to or longer if they choose to and the final schedule allows for this.

To achieve the above mentioned scheduling behaviour, the scheduling process is divided into two steps: The first step creates a provisional schedule, which details which groups sit at which table in which order. The second step further details the provisional schedule and determines the actual start and end time of each group's visit.

To create the provisional schedule the algorithm loops over the demand pattern time slots in order of increasing start times. The start time of a time slot is given by $t_{start;ts}$. For every time slot, the algorithm loops over the number of groups expected to visit the restaurant within this time slot and assigns them to a table according to the logic displayed in Figure 2.

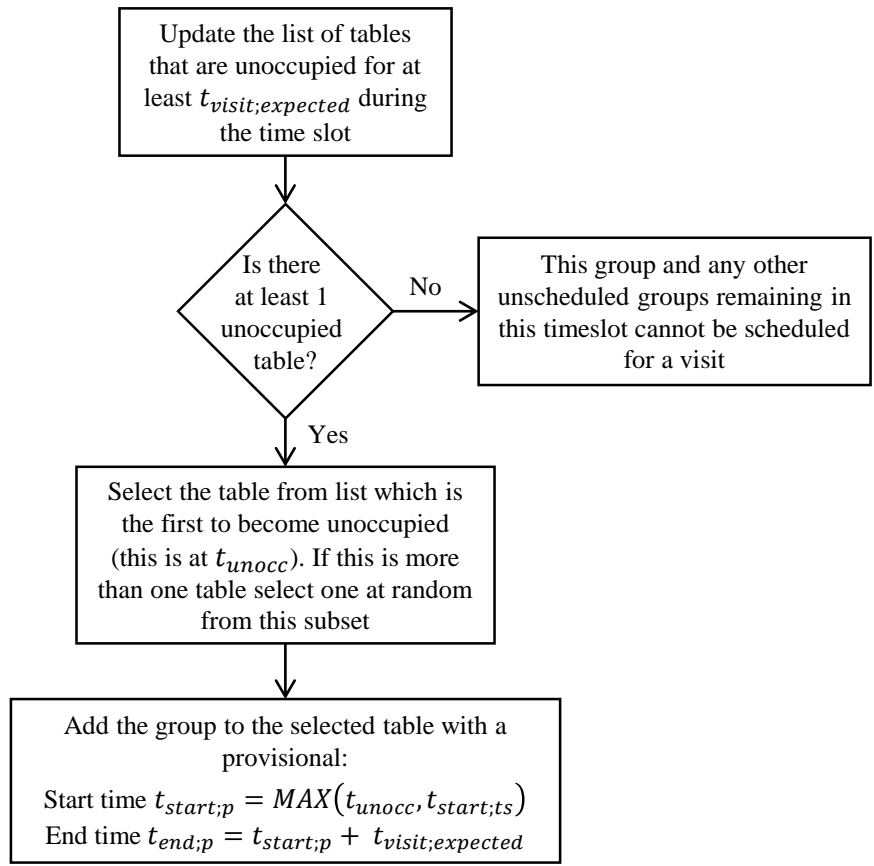

**Figure 2 The algorithm that assigns a group from a given time slot ($t_{start;ts}$ - $t_{end;ts}$) a table and a provisional start and end time for the visit**

The second step sets the actual start and end time of each group's visit by adding some variation in the duration of the visit and the precise start time of the visit. This ensures that the groups within a time slot have slightly different arrival times and visit durations. To achieve this, the algorithm loops over all groups per table in reverse order of arrival. Equations 1-4 are used to calculate the start and end time ($t_{start;i}$ and $t_{end;i}$) for each group.

$$t_{visit;max;i} = MIN(t_{end;ts;i}, t_{end;i+1}) - t_{start;p;i} \quad (1)$$

$$t_{visit;i} = MIN(t_{visit;max;i}, \mathcal{N}(t_{visit;expected}, \sigma_{visit}^2)) \quad (2)$$

$$t_{start;i} = t_{start;p;i} + \mathcal{U}(0, t_{visit;max;i} - t_{visit;i}) \quad (3)$$

$$t_{end;i} = t_{start;i} + t_{visit;i} \quad (4)$$

where

| | | |
|---|---|---|
| $t_{visit;max;i}$ | = | maximal duration of the visit of group $i$ |
| $t_{end;ts;i}$ | = | end time of the time slot |
| $t_{end;i+1}$ | = | end time of the next group occupying the table (is infinite for the last group occupying the table) |
| $t_{start;p;i}$ | = | provisional start time |
| $t_{visit;i}$ | = | duration of the visit |
| $t_{visit;expected}$ | = | the expected visit duration defined in the input |
| $\sigma_{visit}^2$ | = | the variance of the visit duration |

*Individual activity schedule creation*
After all groups have their final start and end times the activity schedules of all individuals can be created. In order to do so, we identified the activities that people perform at a restaurant. The following list of the activities are/can be part of the schedule of an individual visiting a restaurant:
- Enter restaurant: This is always the first activity and a mandatory activity
- Hang coat at coat rack: This is an optional activity performed after entering the restaurant provided a coat rack is available and the customer chooses to use it given an certain probability $P_{coat}$
- Sit at table: A mandatory activity performed after entering the restaurant or using the coat rack
- Go to the toilet: An optional activity provided a toilet is available and the customer chooses to use it given an certain probability $P_{toilet}$
- Pay at the register: A conditional activity assigned to only one member of a group provided the payment is not performed at the table.
- Leave the restaurant: The last activity and a mandatory one.

The list only includes activities related to the customers' movements. Activities at the table, such as ordering food, have no influence on the movement pattern of a customer and are hence not included. They do impact the movements of staff and this is discussed later in the section.
  Figure 3 provides an overview of the algorithm that creates the activity schedule for that individual. The first step is adding the entry activity at the start time of the group. The exact time of entry is computed such that individuals from different groups enter at slightly different times to enable physical distancing. If a coat rack is available, as determined by the input, the individual will potentially hang their coat based on the probability $P_{coat}$. The time needed to perform this activity is given by $D_{coat}$. Next, individuals will walk to the table and sit down at one of the chairs. If a toilet is available, again based on the input, the individual will visit the toilet with a probability equal to $P_{toilet}$. The duration of the toilet visit is in turn determined by the parameter $D_{toilet}$. The toilet visits of all individuals are scheduled such that there are no more toilet visits scheduled at the same time than there are toilets. The underlying assumption is that customers have some indication about how busy the toilets are and plan accordingly. Short queues and waiting times can still occur as the dynamics in the simulation (e.g. walking time) are not taken into account. At the scheduled end time of the group's visit all individuals will move to the exit of the restaurant. If the bill has to be paid at a register one randomly selected individual of the group first moves to the register to pay. The duration of this activity is given by the parameter $D_{register}$. If individuals have their coat at the coat rack, they will first visit the coat rack to collect their coat before moving to the exit.

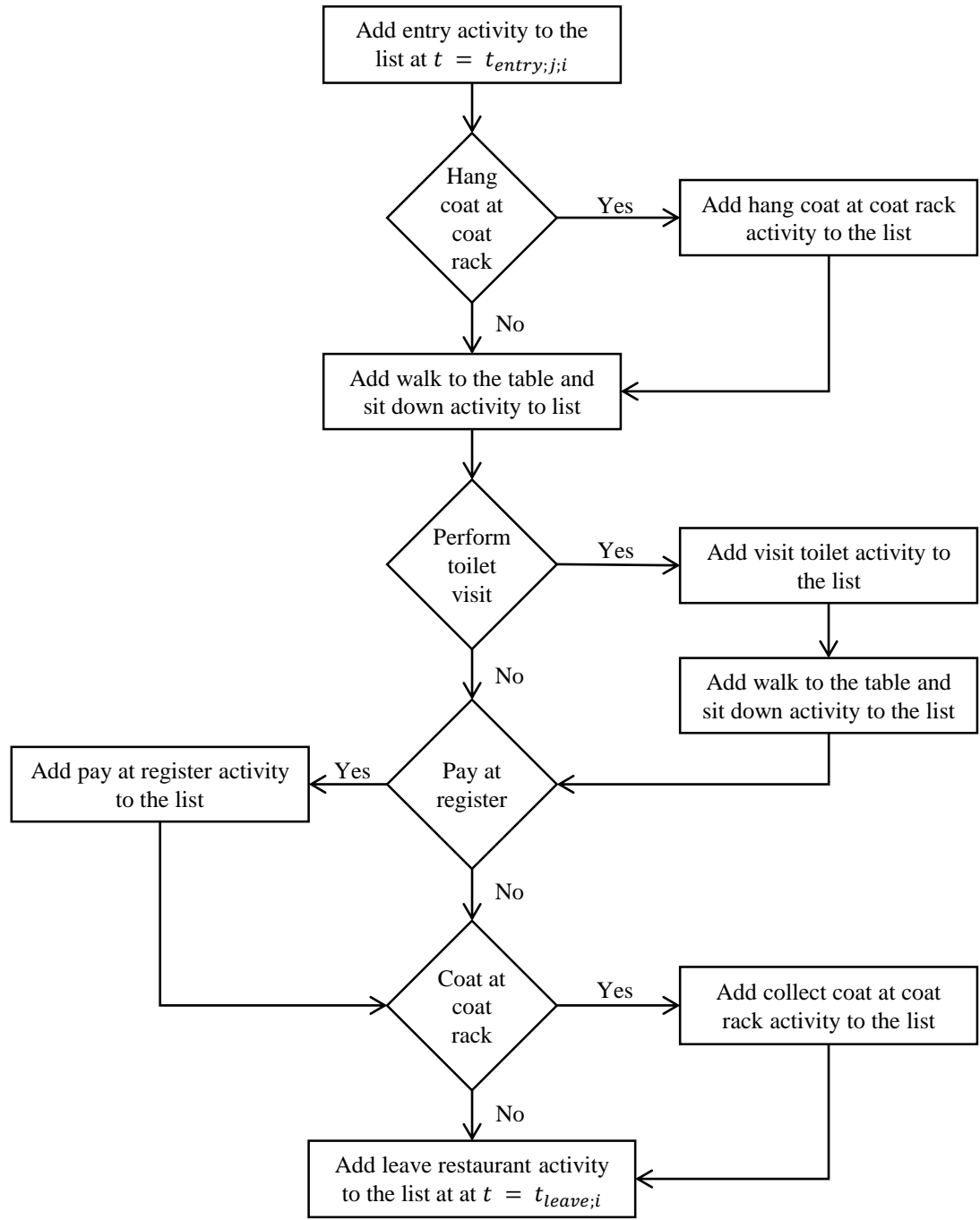

**Figure 3 The activity choice and scheduling algorithm for creating the activity schedule of individual i of group**

## Waiting staff activity choice and scheduling model
The model that chooses and schedules the activities of the waiting staff in the restaurant is an event-based model that dynamically adds activities to be performed and distributes the activities over the staff members.

Again the inputs will be presented first followed by the design of the model.

*Inputs*
The inputs are again chosen such that we only need restaurant owners to provide few and simple inputs. In this case those inputs are:
- The number of waiting staff
- Serving neighbourhoods: A serving neighbourhood is a set of tables that is primarily served by one member of staff
- The average number of times the staff takes and then serves an order at the table
- The staff base area: This is a location in the restaurant where the staff waits if they have no activity to perform and also the location where they collect the food and drink that is to be served.

Also these inputs can be identified by the restaurant owner via a Graphical User Interface of the SSOapp.

*Choice and scheduling model*
The basis of the staff sub-model are activity stacks which contain the information about which activities still need to be performed, when these activities should be performed and in which order. Every neighbourhood has its own stack. There are four events that either add or remove activities from the stack:
1. A group sits down at their table for the first time. This is the only event that adds activities to the stack.
2. A member of staff has finished its current activity and request the next activity from the stack.
3. One or more staff members are available to perform an activity and the start time of one or more activities in the stack is equal to or smaller than the current time.
4. A neighbourhood is overloaded with activities that should be performed and activities are assigned to staff members of other neighbourhood provided they are available to perform an activity.

These four events and their influence on the activity stack are discussed in more detail starting with the event that adds activities to the stack. When a group sits down at their table for the first time, the algorithm determines which activities should be performed by the waiting staff in relation to the visit of this group and when. It first creates the list of activities that should be performed based on the following activities.
- Welcome and distribute menus: This is a mandatory activity that is always the first activity to be performed. It includes collecting the menus at a base location and distributing them at the table and taking the first drinks order.
- Take order: This activity is performed one or more times depending on the number provided by the input. This activity is performed at the table.
- Serve order: For every take order activity a serve order is scheduled whereby the staff member first collects the food or drinks at a base area and then moves to the table to serve it.
- Payment: This activity is only scheduled when customers pay at the table instead of at a register and is the last activity to be scheduled while the customers are still at the table
- Cleaning: This activity is the last activity to be scheduled and is performed after the customers have left the table. After finishing this activity at the table the staff member has to return to the base area to ditch dirty dishes and trash.

Next, it determines for every activity the two properties that control when it can be performed and what its position on the stack will be:

- Earliest start time ($t_{earliest}$): This is the earliest time at which the activity can be started
- Latest start time ($t_{earliest}$): The time before which an activity should ideally be started

The use of the earliest and latest start time properties enable more flexible scheduling of the activities because the activity doesn't have to be performed at exactly one moment in time but can be performed within a certain time frame. For the three activities that are only performed once per visit, equations 5-8 are used to compute their start times.

Welcome activity
$$t_{earliest;w} = t_{latest;w} = t_{sit} \tag{5}$$

Payment activity
$$t_{latest;p} = t_{leave} - d_p - b_{latest;p} \tag{6}$$
$$t_{earliest;p} = t_{latest;p} - b_{earliest;p} \tag{7}$$

Cleaning activity
$$t_{earliest;c} = t_{latest;c} = t_{leave} \tag{8}$$

where
- $t_{sit;j}$ = the time at which the last member of the group sits down at the table
- $t_{leave}$ = the time at which the group leaves the restaurant
- $d_p$ = the duration of the payment activity
- $b_p$ = buffer time parameters

The take order and serve order activities are performed one or more times during the visit and are scheduled to take place between the welcome and payment activities. Equations 9-13 are used to compute the start times for the $k^{th}$ take or serve order activity.

Take order activity
$$t_{earliest;t}(k) = t_{latest;s}(k-1) + d_s + b_s + g_j \tag{9}$$
$$t_{latest;t}(k) = t_{earliest;t}(k) + b_t \tag{10}$$

Serve order activity
$$t_{earliest;s}(k) = \begin{cases} t_{latest;w} + d_w + g_j & \text{If } k = 1 \\ t_{latest;t}(k) + d_t + b_t + g_j & \text{Otherwise} \end{cases} \tag{11}$$
$$t_{latest;s}(k) = t_{earliest;s}(k) + b_s \tag{12}$$

$$g = \frac{1}{2n} \left( t_{earliest;p} - t_{sit} - d_w - n * (d_t + b_t + d_s + b_s) + d_t + b_t \right) \tag{13}$$

where
- $d_s$ = the duration of the serving activity
- $d_w$ = the duration of the welcome activity
- $d_t$ = the duration of the take order activity
- $b_s, b_t$ = buffer time parameters
- $n$ = the number of times a waiter takes and then serves an order as defined by the input

When the activity list is created and all earliest and latest start time are computed the algorithm pushes the list to the activity stack of the correct neighbourhood and sorts the stack in the following order:
1. Latest start time: In ascending order
2. Earliest start time: In ascending order

There are three ways in which staff members can be assigned an activity. The first case is when a staff member has finished its current activity. When this happens the staff member requests the next activity from the stack and the stack return the first activity for which the earliest time is smaller than or equal to the current time. It could be that no activity can be performed yet in which case the staff member is send to the base area to wait for its next activity. The second case is when a staff member has currently no activity to perform and at least one activity on the stack has an earliest time equal to the current time. The staff member is then assigned this activity. Lastly, a free staff member can be assigned an activity from another neighbourhood in case this neighbourhood is overloaded with activities that should be performed now. A neighbourhood is considered to be overloaded with activities if one of the following two conditions hold:
1. There are more than $A_{max;queued}$ activities which should have been performed already (current time is larger than their latest time)
2. There is at least one activity that has been waiting for more than $T_{max;queued}$ to be performed $(t - t_{latest} > T_{max;queued})$.

## SIMULATION EXPERIMENTS AND RESULTS
The previous section presented the two sub-models that enable simulating both the activity schedules of customers and waiting staff in a restaurant. This section puts the model to the test and assesses it. Here, the goal of the simulation experiments is two-fold. Firstly, to assess if the model can indeed be used to assess how the contacts between people in a restaurant change depending on different settings. Secondly, to obtain insight into how activity modelling can assist modelling the risk of virus transmission in a restaurant setting. This section will first describe the setup of the experiment followed by the presentation and analysis of the results.

### Experimental setup
At the time of this study no data was available, so the following experiments are based on a fictitious restaurant (see Figure 4 for its layout). The restaurant has a total of 15 tables and 48 chairs, one toilet, one base area where personnel collects the food and drinks to be served, and a separate entrance and exit to the venue. The layout of the tables and chairs is such that people who are sitting at different tables are at least 1.5 meters apart when sitting or are separated by an airtight barrier.

*Measure of performance*
The measures of performance of interest in this study are the number and duration of contacts between people as these, combined with data about the virus, can provide insight into the likelihood of the virus being transmitted. The contacts between people are captured using a weighted connectivity graph $G$ with weight function $W(\{a_i, a_j\})$.

$$G = \{W(\{a_i, a_j\}) \mid a_i, a_j \in A \text{ and } i \neq j \text{ and } a_i \notin g(a_j) \text{ and } \exists\, t \mid \|\vec{p}_i(t) - \vec{p}_j(t)\| \leq D_{cutoff}\} \quad (14)$$

$$W(\{a_i, a_j\}) = \sum_{t \in T} \Delta t_{\{\|\vec{p}_i(t) - \vec{p}_j(t)\| \leq D_{cutoff}\}}(t) \quad (15)$$

where

$A$ = the set off all agents in the simulation
$g(a_j)$ = the set of agents forming the group to which agent $j$ belongs
$t$ = the time
$T$ = the set of time steps of the simulation
$\Delta t$ = the time step size
$\vec{p}_i(t)$ = the position of agent $i$ at time $t$
$D_{cutoff}$ = the distance cut-off value

This definition shows two things:
1. The weight of a contact is defined as the exposure time which in turn in given by the cumulative time that two people spend within a certain distance of each other, where the threshold distance equals the advised physical distancing distance.
2. We do not consider the contact between members of the same group. That is, contacts between people sitting at the same table and contacts between staff members are ignored, because these are by definition already high risk contacts and would clutter the results whilst providing little additional insights pertaining the functioning of the model.

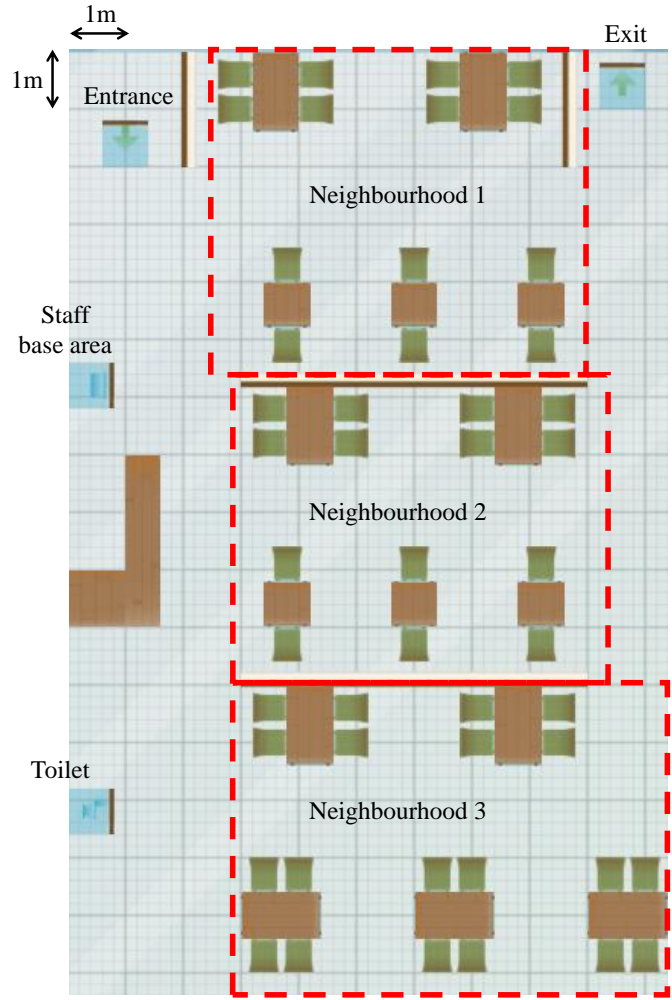

3. **Figure 4 Layout of the restaurant used for the experiments, where the red dashed lines identifies the boundary of the neighbourhoods.**

*Scenarios*
To assess the model, five different scenarios are used. One base scenario and four variations thereof. All five scenarios are discussed in more detail below, whereby the variations will also be compared to the base scenario.

The core properties of the base scenario are that:
1. There are three neighbourhoods with each one member of staff assigned to it.
2. A total of 45 groups will visit the restaurant in three separate time slots (so 15 groups per time slot) whereby there is a short period of time between the time slots during which the restaurant is empty. This is one of possible the measures that can prevent contacts between different groups visiting the restaurant.
3. Each group has an average visit duration of 1.5 hours during which they get 4 servings of drinks and food.

The variations are chosen to represent different restaurant settings. This enables us to assess how the model exactly influences the contacts between people in these different settings. Furthermore, as we can also reason what changes we would expect when comparing (qualitatively) a variation to the base scenario, we can also verify if the model behaves as expected. The four variations are described below including how we expect the results to change compared to the base scenario.

The first variation is the 'No time slots' scenario. In this scenario there are no time slots and the 45 groups will visit in a more randomly fashion. All other inputs and parameters are the same as the base scenario. The expected result is that there are more simultaneous movements in the restaurant and hence more and/or longer contacts between people from different groups.

The second variation is the 'Longer visits' scenario. In this scenario fewer groups, in this case 30, visit the restaurant but they stay for a longer period of time, 2 hours and 20 minutes instead of 1.5 hours, have more servings, 7 instead of 4 and have a higher likelihood of visiting the toilet. The expected result is that there are fewer contacts per person but that these contacts are longer in duration.

The third variation is the 'No neighbourhoods' scenario. In this case the three staff members aren't assigned to any neighbourhood but can serve all tables. The expected result is that staff members come in contact with more people but that these contacts are of a shorter duration.

The last variation is the 'More toilet visits' scenario where all guests visit the toilet once during their visit. The expected result is that there are more movement within the restaurants and thus more contacts.

Details regarding the inputs and parameters for all five scenarios are found in Table 1 and Table 2. The first table present the inputs that differ between the four scenarios whilst the second table presents the inputs and parameters shared by all five scenario. For all five scenarios, the weighted connectivity graph was obtained using approximately 500 replications for every scenario. A cut-off distance of 1.5 meters was used as this is a commonly advised physical distancing distance.

**Table 1  Overview of the inputs that differ between the scenarios**

|  | Slots | Groups visiting [#] | Visit duration [min] | P toilet [-] | Order count [#] | Neighbour-hoods |
|---|---|---|---|---|---|---|
| **Base** | Yes | 45 | $\mathcal{N}(90,5)$ | 0.4 | 4 | Yes |
| **No time slots** | No | 45 | $\mathcal{N}(90,5)$ | 0.4 | 4 | Yes |
| **Longer visits** | Yes | 30 | $\mathcal{N}(140,5)$ | 0.8 | 7 | Yes |
| **No neighbourhoods** | Yes | 45 | $\mathcal{N}(90,5)$ | 0.4 | 4 | No |
| **More toilet visits** | Yes | 45 | $\mathcal{N}(90,5)$ | 1 | 4 | Yes |

**Table 2  Overview of the inputs and parameters shared by all four scenarios**

| Name | Value |
|---|---|
| **Inputs** | |
| Personnel count [#] | 3 |
| Toilet visit duration [s] | $\mathcal{N}(120, 30)$ |
| | |
| **Parameters** | |
| Welcome activity duration ($d_w$) [s] | 120 |
| Payment activity duration ($d_p$) [s] | 60 |
| Cleaning activity duration ($d_c$) [s] | 150 |
| Take order activity duration ($d_t$) [s] | 45 |
| Serve order activity duration ($d_s$) [s] | 30 |
| Buffer time payment latest ($b_{latest;p}$) [s] | 120 |
| Buffer time payment earliest ($b_{earliest;p}$) [s] | 120 |
| Buffer time take order ($b_t$) [s] | 120 |
| Buffer time serve order ($b_s$) [s] | 120 |

**Results**

Two types of analyses are used to determine the face validity of the new model. The aim of these analyses is to determine whether the model functions as expected. The first analysis focusses on the differences between the scenarios. A second analysis will investigate if and how the cut-off distances impact the results.

Figure 5 presents the cumulative distribution functions of the exposure times over the contacts for all five scenarios. The first observation is that the overwhelming majority (~80%) of the contacts are of short duration (<20 seconds). The insets in the graph show that these short duration contacts include almost all of the contacts between customers and more than half of the contacts between the staff and customers. These short contacts are most likely the results of agents passing each other one or more times. The insets also show that the differences between the scenarios are mainly caused by the duration of the staff-customer interactions. This result is unsurprising, given that staff stands next to the table whilst taking or serving an order, which takes longer than simply passing whilst walking.

The 'No time slots' scenario and the 'More toilet visits' scenario both have a limited impact on the distribution of the exposure time over the contacts compared to the base scenario. However, Figure 6 (a) and (b) show that in both scenarios customers have slightly more contact with each other and also with the restaurant staff. This is in line with our expectation, because the lack of time slots and the increased toilet visits raise the amount of simultaneous movements in the restaurant. The increased toilet visits, however, do also impact the amounts of contacts staff has with customers (see Figure 6 (c)) whilst this is not the case when there are no time slots. This, again is logical as the increased toilet visits increase the overall number of movements whilst the lack of time slots only increases the likelihood of simultaneous movements.

The 'No neighbourhoods' and 'Longer visits' scenarios clearly lead to bigger differences in the contacts between staff and customers in comparison to the base scenario. The 'No neighbourhoods' scenario results in shorter contact times between staff and customers (Figure 5) but more contacts per person compared to the base scenario (Figure 6 (c)). Furthermore, regarding the customer-customer contacts there are no statistically significant differences to the base scenario. This is exactly as expected, as the lack of neighbourhoods should indeed only affect the staff-customer interactions and should show a negative correlation between the amount of contacts and the duration of the contacts for the staff-customer interactions.

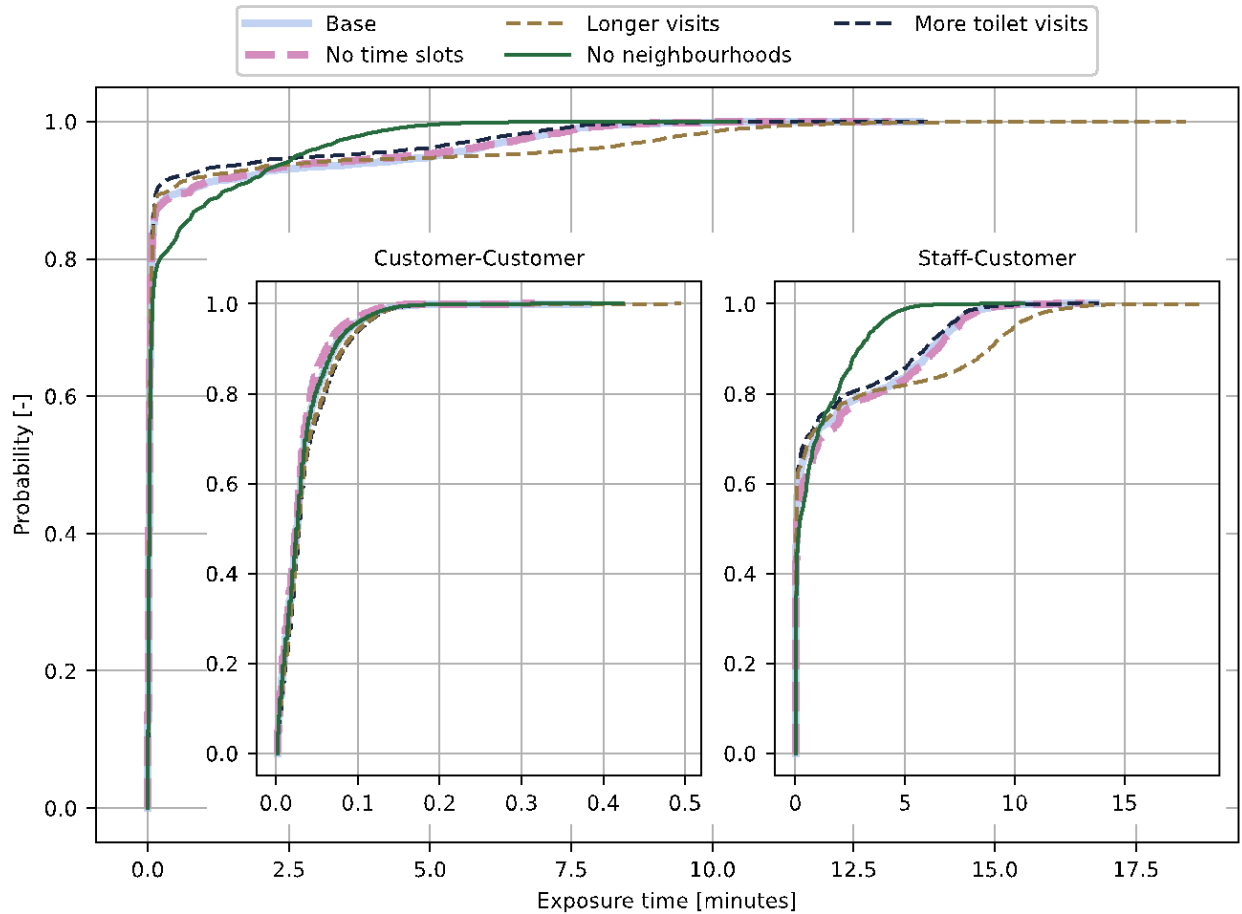

**Figure 5  The empirical CDF of the weight over the contacts for all four scenarios**

The 'Longer visits' scenario, moreover, results in longer contacts on average between customers and staff (Figure 5). However, it also results in fewer contacts between staff and customers (Figure 6 (c)). The longer contacts are a logical consequence of more servings and, hence, more table visits by the staff and the fact that there are fewer contacts is also logical as fewer people visit the restaurant. Due to the fact that there are also more toilet visits, customers are a bit more likely to come in contact with one another and with more than one member of staff than is the case for the base scenario (Figure 6 (b) and (c)).

**Sensitivity**
The distance threshold adopted to compute the number of risky contacts is the only parameter in the analysis that has no direct relation to the restaurant setting. Furthermore, the distance that is advised for physical distancing also differs per country and might deviate from the 1.5m used above. Hence, we want to obtain insight into how sensitive the outcomes are to a change in this distance.
    Analysing the same cdfs for 1.0m and 2.0m cut-off distances shows that distance has an influence on both the weight of the contacts and the number of contacts per agent. For the 1.0m case the weights are lower and there are fewer contacts compared to the 1.5m case and for the 2.0m case it is vice versa. This is also as can be expected. The effect is biggest for the customer-customer interactions which seems to indicate that the sensitivity to the cut-off distance is mainly related to the movements through the restaurant and not so much to the interaction between customers and staff at the table. The differences in the patterns between the different scenarios however remain. So, the choice of cut-off distance does impact the results however the relative differences between the different settings are not impacted.

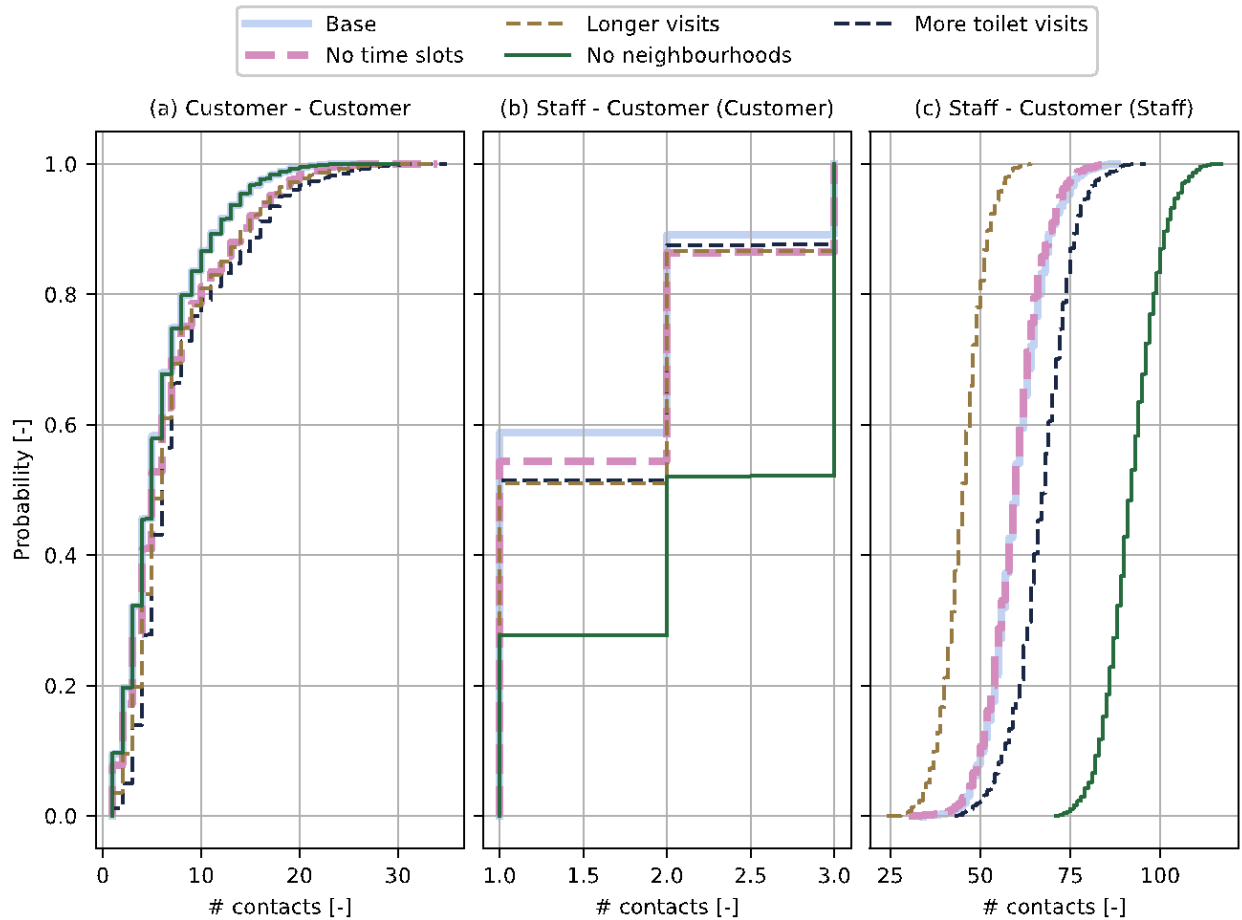

**Figure 6** The empirical CDFs of the # contacts over the agents for the different types of interactions and agents

## CONCLUSIONS, DISCUSSIONS AND IMPLICATIONS
With the advent of the SARS-CoV-2 virus, a new challenge arrived for the field of pedestrian dynamics. High density situations, one of the prime interests, became scares and interest moved to the question of how pedestrian behaviour modelling can assist in assessing the risk of virus transmission. One location of interest is restaurants where the question is how a restaurant can be set up to reduce the risk of virus transmission. This posed a challenge as models to create activity schedules for restaurant settings have had very little attention and as such none were available for use.

This paper introduced a novel activity choice and scheduling model which focusses on producing realistic activity schedules in a restaurant setting for both customers and waiting staff. The main feature of the model is that is requires only few inputs but still can model different restaurant settings. This is important as the model is intended to be used by restaurant owners who have neither the time nor the experience to work with complex models. Tests of the model show that the model can produce face valid activity schedules for both customer and waiting staff. More importantly, the tests do also show that the contact patterns found are consistent with what would be expected and that these also change as expected when the restaurant setting changes.

The adoption of the activity scheduler (and the SSOapp), are important when one is interested in obtaining insight into how different measures in a restaurant influence the contacts between people. This last part is especially relevant in the current day and age. When combined with data about the virus, this contact data can help inform which measures can help in preventing the spread of an airborne virus, such as the SARS-CoV-2 virus. For example, if we take the case studied in this paper, it is clear the staff are the biggest risk factor for the spreading of the virus whilst simultaneously they are also at the biggest risk of catching it. This insight could inform that, for example, a policy of very regular testing of restaurant staff is indeed a wise thing to do and that measures, other than physical distancing, such as face masks for the staff, might also be relevant to consider.

However, this observation is rather straightforward and one might not need a model to reason that this is the case. However, these sort of results can also be used to obtain much less obvious insights. For example, what setting, neighbourhoods versus no neighbourhoods or more visitors with shorter visiting durations versus fewer visitors with longer visiting durations poses less risk of virus transmission? And how is this dependent on how virulent a particular virus or strain is? Take the case of a very virulent strain where only a short cumulative contact duration is necessary to be at high risk for transmission. In this case, using neighbourhoods decreases the transmission risk as it decreases the number of contacts. The same hold for fewer visitors, even if this means longer visit durations. On the other hand, if a strain is less virulent, the cumulative contact time becomes more relevant and using no neighbourhoods can actually reduce the risks as the probability that a contact will exceed a certain time is lower.

This study also has some limitations. Firstly, due to the lack of data at the time of this study no validation could be performed and the evaluation of the model's performance was limited to face validation. Hence, our plans are to collect activity schedule data in restaurants as soon as the restrictions allow for it. Secondly, the tests were limited to one restaurant layout. This means that the question of generalizability is still an open one.

Nonetheless, the results show the importance of activity scheduling models in situations such as restaurants where far fewer movements occur than in, for example, a busy transport hub and hence the activity scheduling has a bigger impact on the results. They also show the potential of these kinds of activity scheduling models when they are combined with information about a virus. When combined, these can inform policy makers and restaurant owners about which strategies can reduce the risks posed by the virus and how this depends on the properties of the virus.

Overall, this study shows that pedestrian dynamics modelling can be of high relevance in these pandemic times. And more importantly, that due to the fact that different dynamics and situations are of interest that activity choice and scheduling modelling is more important and relevant than ever.

## ACKNOWLEDGEMENTS
This publication is part of the project SamenSlimOpen (with project number 10430022010018 of the research programme COVID-19 Programma, which is financed by the Dutch Research Council (NWO) and ZonMw. We thank Quirine Ten Bosch, Büsra Atamer Balkan, Colin Teberg, Berend Wouda, Doris Boschma, Linda van Veen, Marjon Koopmans and Wim van der Poel for their active involvement & fruitfull discussions in the SamenSlimOpen.

## AUTHOR CONTRIBUTIONS
The authors confirm contribution to the paper as follows: study conception and design: M. Sparnaaij, D. Duives, Y. Yuan; simulation coding and execution: M. Sparnaaij; analysis and interpretation of results: M. Sparnaaij; draft manuscript preparation: M. Sparnaaij, D. Duives, Y. Yuan, W. Daamen. All authors reviewed the results and approved the final version of the manuscript.